\documentclass[twocolumn,showpacs,preprintnumbers,amsmath,amssymb]{revtex4}
\usepackage{graphicx}
\usepackage{dcolumn}
\usepackage{bm}
\usepackage{amsfonts}
\usepackage{amsmath}
\usepackage{amssymb}

\begin{document}

\title{\Large \bf Hyperdeformation in the Cd isotopes: a microscopic analysis}

\author{ H.\ Abusara and  A.\ V.\ Afanasjev }

\affiliation{Department of Physics and Astronomy, Mississippi State
University, MS 39762, USA} 

\date{\today}

\begin{abstract}
 A systematics search for the nuclei in which the observation of 
discrete hyperdeformed (HD) bands may be feasible with existing 
detector facilities  has been performed in the Cd isotopes within 
the framework of cranked relativistic mean field theory. It was 
found that the $^{96}$Cd nucleus is a doubly magic HD nucleus due 
to large proton $Z=48$ and neutron $N=48$ HD shell gaps. The best 
candidate for experimental search of discrete HD bands is $^{107}$Cd 
nucleus characterized by the large energy gap between the yrast and 
excited HD bands, the size of which is only 15\% smaller than the 
one in doubly magic HD $^{96}$Cd nucleus.
\end{abstract}


\maketitle

\section{Introduction}

Hyperdeformation (HD) is one of critical phenomena in nuclear structure, 
the study of which will considerably advance our knowledge of nuclei at 
extreme conditions of very large 
deformation and fast rotation \cite{CRMF-HD,Dudek}. The studies of HD will 
also contribute into understanding of the crust of neutron stars, where 
extremely deformed nuclear structures are expected  (see Ref. \cite{stars}
and references therein). Although some experimental evidences of the 
existence of HD at low \cite{K.98,C12} and high spin 
\cite{152Dy-exp-1,152Dy-exp-2,Hetal.06,60Zn-1}  exist, the current experimental 
knowledge of HD is very limited. New generation of detectors such as 
GRETA \cite{Greta} and AGATA \cite{Agata} will definitely allow to study 
this phenomenon in more details. However, these detectors will  become 
functional only in the middle of  next decade. Thus, it is very important 
to understand whether new experimental information on HD can be obtained 
with existing detectors such as GAMMASPHERE \cite{Gamma}.

Theoretical efforts to study HD at high spin both in macroscopic+microscopic 
(MM) method and in self-consistent approaches were reviewed in Ref.\ \cite{CRMF-HD}. 
Our recent  study of HD within the framework of the cranked relativistic mean field 
(CRMF) theory in the Z=40-58 part of nuclear chart 
\cite{CRMF-HD} represents the first ever systematic investigation of HD within 
the self-consistent theory. The general features of the HD bands at high spin have 
been analysed in Ref. \cite{CRMF-HD}. 
In particular, it was concluded that the density of the HD states in the vicinity of 
the yrast line is the major factor which decides whether or not discrete HD bands can 
be observed. The high density of near-yrast HD states will lead to a 
situation in which the feeding intensity will be redistributed among many 
low-lying bands, thus drastically reducing the intensity with which each 
individual band is populated. For such densities, the feeding intensity of an 
individual band will most likely drop below the observational limit of the 
modern experimental facilities. On the contrary, the large energy gap between 
the yrast and excited HD configurations will lead to an increased population of 
the yrast HD band, thus increasing the chances of its observation.

  The analysis of Ref. \cite{CRMF-HD}, based on the energy gap between the last 
occupied and first unoccupied routhians in the yrast HD configurations, 
suggests that the density of the HD bands in the spin range where they are 
yrast is high in the majority of the cases. It also indicates the Cd isotopes as 
the best candidates for a search of discrete HD bands. However, one has to 
remember that this type of analysis may be too simplistic
because the polarization effects induced by particle-hole excitations are 
neglected. In particular, it can overestimate the size of the 
energy gap between the yrast and excited HD configurations. Realistic analysis of
the density of the HD bands should include significant number of the HD configurations 
calculated in a fully self-consistent manner with all polarization effects 
included. Such analysis is time-consuming in computational sense and has been
performed only for $^{124}$Xe in Ref. \cite{CRMF-HD}, but its extension to
other nuclei is needed. Thus, the goals of the current manuscript are (i) 
to perform a fully self-consistent analysis of the density of the HD bands in 
the Cd isotopes and (ii) to find the best nuclei in which experimental study of 
discrete HD bands can be feasible with existing experimental facilities.

\section{Theoretical framework and the details of the calculations}

   The calculations in the present manuscript are performed in the framework 
of the CRMF  theory without pairing \cite{KR.89,A150} using numerical scheme
of Ref.\ \cite{CRMF-HD}. The CRMF equations for the HD states are solved in the 
basis of an anisotropic three-dimensional harmonic oscillator in Cartesian 
coordinates characterized by the deformation parameters $\beta_0=1.0$ and 
$\gamma=0^{\circ}$ and oscillator 
frequency $\hbar \omega_0= 41 A^{-1/3}$ MeV (see Ref.\ \cite{CRMF-HD} for details).
The truncation of basis is performed in such a way that all states belonging
to the shells up to fermionic  $N_F$=14 and bosonic $N_B$=20 are taken 
into account;
this truncation scheme provides sufficient numerical accuracy \cite{CRMF-HD}.
The NL1 parametrization of the RMF Lagrangian \cite{RRM.86} is used
in most of our calculations since it provides a good description of the moments
of inertia of the rotational bands in unpaired regime in the SD and ND minima 
\cite{A150,ALR.98,A60,VRAL.05}, the single-particle energies for the
nuclei around the valley of $\beta$ stability \cite{ALR.98,A250} and the
excitation energies of the SD minima \cite{LR.98}. Other parametrizations 
such as NL3 \cite{NL3}, NLSH \cite{NLSH}, NLZ \cite{NLZ} and NL3* 
\cite{NL3*} are used only to check the size of the HD gaps in the nuclei of 
interest. 

\begin{figure}[ht]
\includegraphics[angle=0,width=6.3cm]{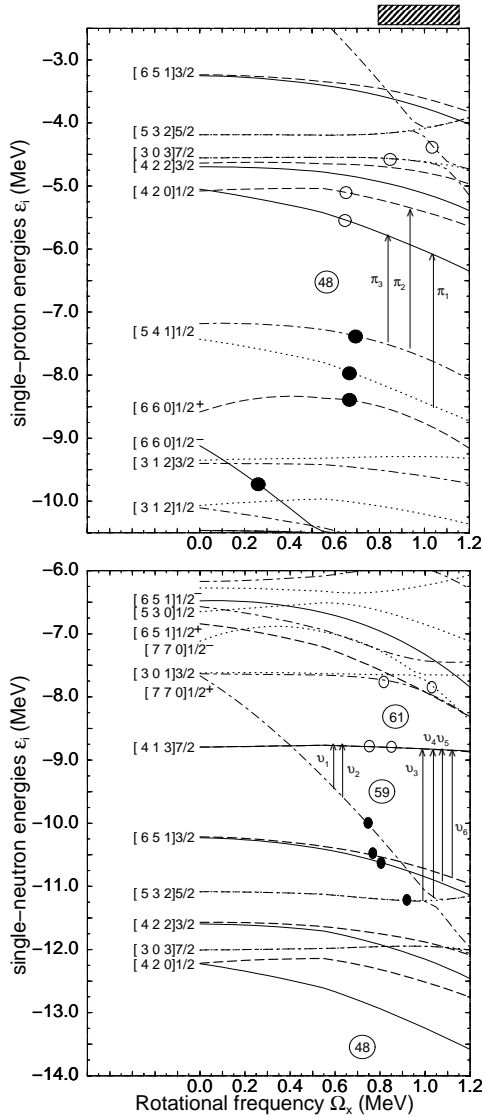}
\vspace{0.0cm}
\caption{Proton (top panel) and neutron (bottom panel) single-particle 
energies (routhians) in the self-consistent rotating potential as a 
function of the rotational frequency $\Omega_x$. They are given along 
the deformation path of the yrast HD configuration 
in $^{107}$Cd and obtained in the calculations with the NL1 parametrization 
of the RMF Lagrangian. Long-dashed, solid, dot-dashed and dotted lines 
indicate  $(\pi=+, r=+i)$, $(\pi=+, r=-i)$, $(\pi=-, r=+i)$ and 
$(\pi=-, r=-i)$ orbitals, respectively. Solid (open) circles indicate the 
orbitals occupied (emptied). The dashed box indicates the frequency range 
corresponding to the spin-range $I=55-80\hbar$ in this configuration. The 
arrows indicate the particle-hole excitations leading to excited HD 
configurations.}
\label{Cd107-both}
\end{figure}

  Single-particle orbitals are labeled by $[Nn_z\Lambda]\Omega^{sign}$.
$[Nn_z\Lambda]\Omega$ are the asymptotic quantum numbers (Nilsson quantum 
numbers) of the dominant component of the wave function at $\Omega_x=0.0$ MeV. 
The superscripts {\it sign} to the orbital labels are used sometimes to 
indicate the sign of the signature $r$ for that orbital $(r=\pm i)$.

 The excited HD configurations were built from the yrast HD configurations 
obtained in the previous study \cite{CRMF-HD}
by exciting either one proton or one neutron or both together. 
Proton and neutron configurations generated in this way are labeled by 
$\pi$$_{i}$ and $\nu$$_{j}$, where $i=0,1,2,...$ and $j=0,1,2,...$ are integers 
indicating the 
corresponding configurations. $\pi$$_{0}$ $\otimes$ $\nu$$_{0}$ represents the 
yrast HD configuration. Total excited configurations $\pi$$_{i}$ $\otimes$ 
$\nu$$_{j}$ are constructed from all possible combinations of proton 
$\pi$$_{i}$ and neutron $\nu$$_{j}$ configurations excluding the one with 
$i=0$ and $j=0$. The selection of excited configurations is also constrained by 
the condition that the energy gap between the orbital from which the particle is 
excited and the orbital into which it is excited do not exceed 2.5 MeV in the 
routhian diagram for the yrast HD configuration.  All configurations are 
calculated in a fully self-consistent manner so that their total energies are 
defined as a function of spin.

Fig. \ref{Cd107-both} illustrates the selection of excited configurations. It 
shows the occupation of the proton and neutron orbitals in the yrast HD 
configuration in $^{107}$Cd. According to our criteria only three proton 
excitations across the $Z=48$ HD gap are considered. On the contrary, more 
neutron $ph$-excitations are allowed across the $N=59$ HD shell gap. 
Table \ref{Table1} shows their detailed structure. For example, the 
$\nu$$_{1}$ configuration is created by exciting one neutron from the 
[770]1/2$^{+}$ into [413]7/2$^{+}$ orbitals. One can notice that we only consider the 
$ph$-excitations between the states which do not have the same combination ($\pi$,$r$)
of parity $\pi$ and signature $r$. The computer code in general can 
handle the excitations between the states with the same ($\pi$, $r$), but the
configurations based on such excitations are less numerically stable and
require more computational time. 
Because of this reason and the fact that they do not alter significantly the 
results for the density of the HD states, it was decided to neglect them in
the calculations. However, in the cases of large energy gaps between the yrast 
and excited HD configurations, they are taken into account.

\begin{table}[ht]
\vspace{-0.5cm}
\caption{Neutron particle-hole excitations in $^{107}$Cd shown in Fig.
\ref{Cd107-both}.}
\newcommand{\m}{\hphantom{$-$}}
\newcommand{\cc}[1]{\multicolumn{1}{c}{#1}}
\renewcommand{\tabcolsep}{0.5pc} 
\renewcommand{\arraystretch}{1.4} 
\begin{tabular}{ccc} \hline
 &  label          & Excitation  \\   \hline
 &  $\nu$$_{1}$    & [770]1/2$^{+}$ $\rightarrow$ [413]7/2$^{+}$ \\     
 &  $\nu$$_{2}$    & [770]1/2$^{+}$ $\rightarrow$ [413]7/2$^{-}$ \\      
 &  $\nu$$_{3}$    & [532]3/2$^{-}$ $\rightarrow$ [413]7/2$^{+}$  \\     
 &  $\nu$$_{4}$    & [532]3/2$^{-}$ $\rightarrow$ [413]7/2$^{-}$  \\ 
 &  $\nu$$_{5}$    & [651]3/2$^{-}$ $\rightarrow$ [413]7/2$^{+}$ \\      
 &  $\nu$$_{6}$    & [651]3/2$^{+}$ $\rightarrow$ [413]7/2$^{-}$ \\ \hline     
\end{tabular}\\[2pt]
\label{Table1}
\end{table}

\begin{figure*}[t]
\vspace{-1.2cm}
\includegraphics[angle=0,width=16.0cm]{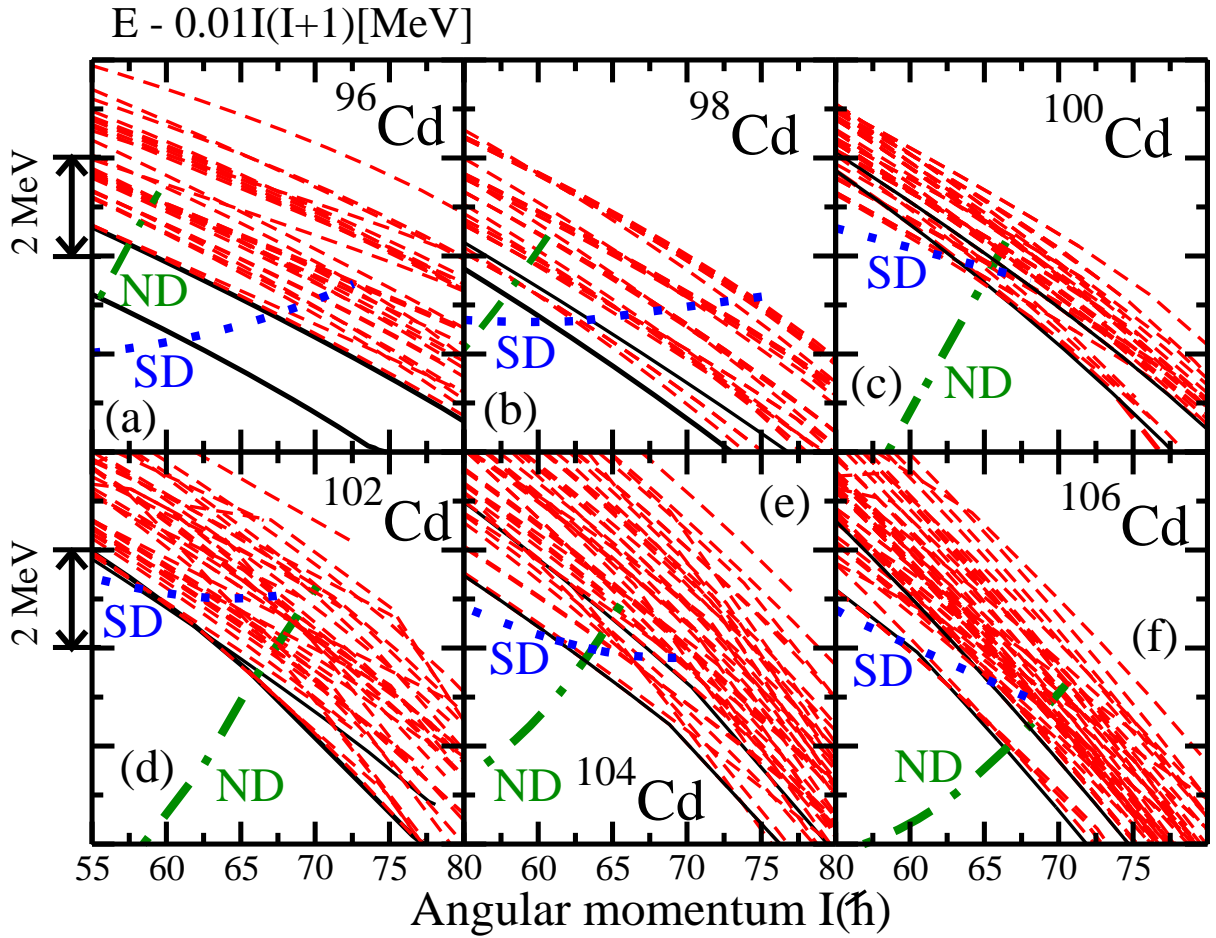}
\vspace{0.50cm}
\caption{(Color online) Energies of the calculated HD configurations in 
even-even $^{96-106}$Cd nuclei relative to a smooth liquid drop reference
$AI(I+1)$, with the inertia parameter A=0.01. In each nucleus, the yrast 
and lowest excited proton configurations are shown by solid lines.
Dot-dashed and dotted lines represent the yrast lines at low spin built from 
normal-deformed (ND) and superdeformed (SD) states, respectively.}
\label{DOS-HD}
\end{figure*}

\section{Discussion}

\begin{figure*}[ht]
\includegraphics[angle=0,width=16.0cm]{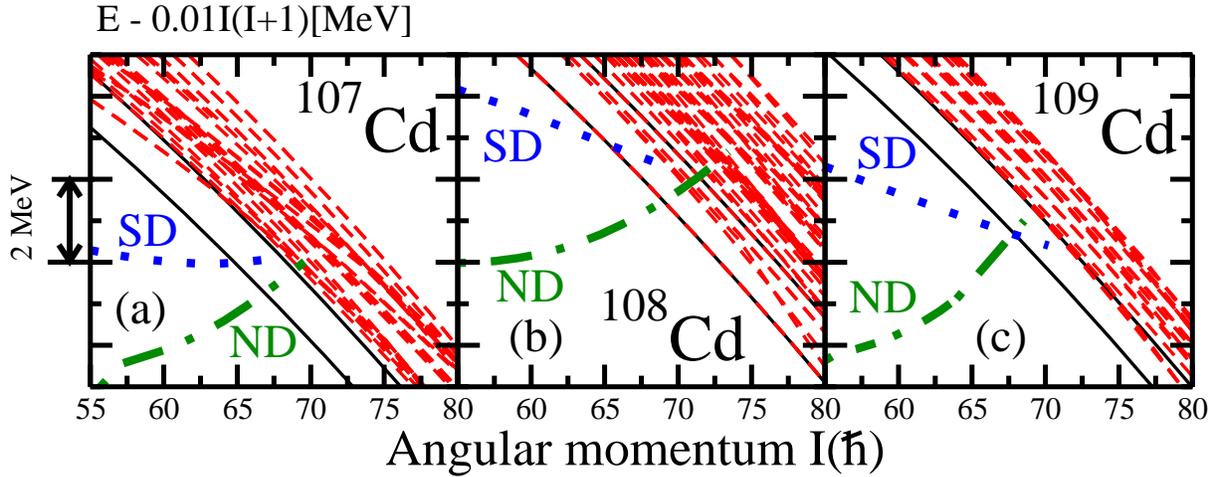}
\vspace{3.50cm}
\caption{ (Color online) The same as in Fig. \ref{DOS-HD} but for 
$^{107,108,109}$Cd. The yrast HD line in $^{108}$Cd is built from two 
signature-degenerated configurations.}
\label{DOS-HD-odd}
\end{figure*}

  Figs.\ \ref{DOS-HD} and \ref{DOS-HD-odd} show the density of the HD states 
in even-even $^{96-108}$Cd and odd mass $^{107,109}$Cd nuclei studied using 
above outlined procedure. The energy gap between the yrast HD configuration and 
lowest excited HD configurations is around 1.5 MeV in $^{96}$Cd (Fig.\ 
\ref{DOS-HD}a). It is comparable with the energy gap between the yrast
and excited SD configurations in doubly magic SD nucleus $^{152}$Dy 
(Fig.\ 7 in Ref.\ \cite{A150}). This energy gap in $^{96}$Cd is due to large 
energy cost of particle-hole excitations across the $Z=48$ and $N=48$ HD 
shell gaps which have similar size (see Fig.\ \ref{Cd107-both} and Table 
\ref{Table2}). All that together indicates that the {\it $^{96}$Cd is a 
doubly magic HD nucleus.} Only proton excitations to the $[420]1/2^-$ orbital 
above the $Z=48$ HD shell gap result in bound excited proton configurations, 
the excitations to other orbitals located above the $Z=48$ HD shell gap 
produce the proton-emitting states. The doubly magic nature of $^{96}$Cd nucleus 
is confirmed also in the calculations with other RMF parametrizations (Table 
\ref{Table2}). It is interesting to mention that the RMF parametrizations aimed 
at the description of the nuclei far from stability such as NL3, NL3*, NLSH 
show larger $Z=48$ and $N=48$ HD shells gaps in $^{96,107-109}$Cd than the 
parametrizations NL1 and NLZ fitted predominantly to $\beta$-stability 
nuclei (Table \ref{Table2}).

  With increasing neutron number the energy gap between the yrast and 
excited HD configurations disappears (Fig.\ \ref{DOS-HD}). This is due 
to relatively high density of the neutron states above the $N=48$ HD 
shell gap (Fig.\ \ref{Cd107-both}).
Indeed, many excited neutron configurations are located below the lowest 
excited proton configurations (Fig.\ \ref{DOS-HD}). One can also see that 
even-even $^{100-104}$Cd nuclei are characterized by appreciable density of 
the HD states in the vicinity of the yrast HD line
(Fig.\ \ref{DOS-HD}). The analysis of the single-particle 
structure in these nuclei indicates that similar density of the HD bands
is expected also in odd mass nuclei $^{99-105}$Cd. In no way these nuclei 
have to be considered as good candidates for a search of discrete HD 
bands since the feeding intensity will be redistributed among many 
low-lying HD bands.  As a result, the feeding intensity of an individual HD 
band will most likely drop below the observational limit of modern 
experimental facilities. Although there is some energy gap between the lowest 
four HD configurations  and other excited configurations in $^{106}$Cd, this 
nucleus does not appear to be a good candidate for a search of discrete HD 
bands because the presence of four low-lying HD configurations will lead to 
a fragmentation of feeding intensity. This is one of possible reasons why the 
HD bands have not been observed in this nucleus \cite{F.05}.

 On the other hand, the high density of the HD bands in above discussed nuclei 
will most likely favor the observation of the rotational patterns in the form of 
ridge structures in three-dimensional rotational mapped spectra \cite{Hetal.06}.
The study of these patterns as a function of neutron number can provide a valuable 
information about HD at high spin.

\begin{figure*}[ht]
\includegraphics[angle=0,width=16.0cm]{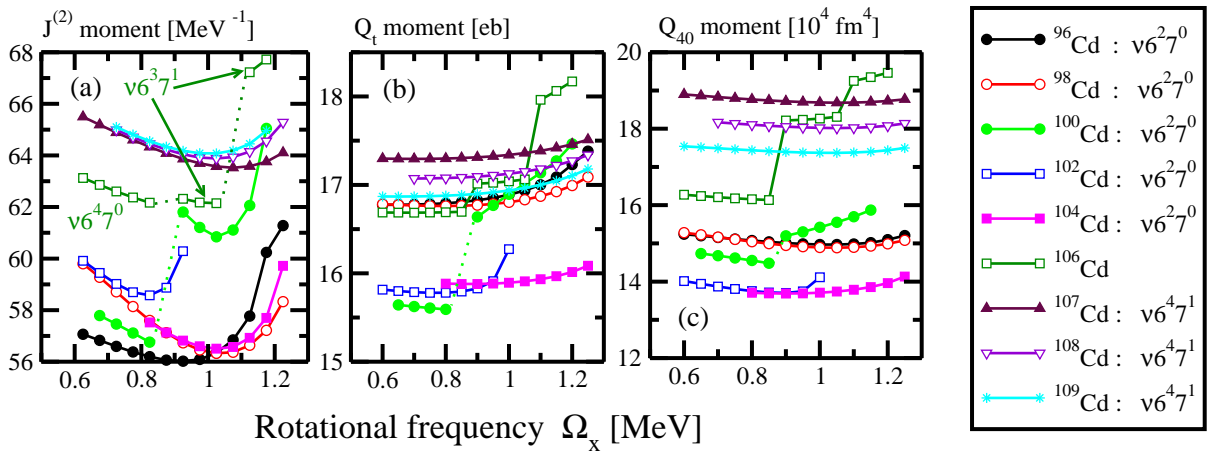}
\vspace{1.00cm}
\caption{ (Color online) Dymanic moments of inertia $J^{(2)}$ (panel (a)), 
transition quadrupole moments (panel (b)), and mass hexadecapole moments
$Q_{40}$ (panel (c)) of the yrast HD bands in the nuclei under 
study. All these bands have the proton $\pi 6^2$ configuration, their 
neutron configurations are shown in the right panel. The configuration
of the yrast HD band in $^{106}$Cd is shown in panel (a).}
\label{Calc-quant}
\end{figure*}

  Further increase of the neutron number brings the neutron Fermi level to 
the region of low density of the neutron states characterized by the 
large $N=59$ and $N=61$ HD shell gaps (Fig.\ \ref{Cd107-both}) with the 
combined size of these two gaps being around 2.5 MeV (Table \ref{Table2}).
As a result, the $^{107-109}$Cd nuclei show appreciable energy gap between 
the yrast and lowest excited HD configurations (Fig.\ \ref{DOS-HD-odd}).
This gap is especially pronounced in the case of $^{107}$Cd for which it 
is around 1.3 MeV. Note that the size of this gap is defined by the size 
of the $Z=48$ HD shell gap, since the lowest excited configuration is 
based on proton excitation (Fig.\ \ref{DOS-HD-odd}a). Similar or even larger 
energy gap between the yrast and excited HD configurations is expected in 
the NLZ, NL3, NL3* and NLSH parametrizations for which the size of the $Z=48$ 
and $N=59$ HD shell gaps is at least 1.7 MeV in $^{107}$Cd (Table \ref{Table2}).  
The energy gaps between the yrast and excited HD configurations at the spins 
where the HD configurations become yrast are somewhat lower in $^{108,109}$Cd 
being around 0.9 and 1.1 MeV. This energy gap in $^{108}$Cd is dictated by the 
size of the $N=61$ HD shell gap since lowest excited HD configurations are 
based on neutron excitations. Thus, in $^{108}$Cd it will be smaller (similar) 
in the case of the NLZ (NL3, NL3*) parametrizations and larger in the NLSH 
parametrization as compared with the one obtained in the NL1 parametrization 
(Table \ref{Table2}). In the case of $^{109}$Cd, the energy gap between the 
yrast and excited configurations will be larger (smaller) in the NL3, NL3* 
and NLSH  (NLZ) parametrizations (Table \ref{Table2}).

\begin{table}[ht]
\vspace{-0.2cm}
\caption{The size of the Z=48, N=59, and N=61 HD shell gaps (in MeV) 
obtained with different parametrizations of the RMF Lagrangian for the 
yrast HD configurations in $^{96,107,108,109}$Cd. They are given at 
rotational frequency $\Omega_x=1.00$ MeV approximately corresponding to 
the spin at which the HD bands become yrast. The lowest (among the 
different parametrizations) value of the shell gap is shown by bold 
style. '59+61' line shows the combined size of the $N=59$ and $N=61$
HD shell gaps.}
\newcommand{\m}{\hphantom{$-$}}
\newcommand{\cc}[1]{\multicolumn{1}{c}{#1}} 
\renewcommand{\tabcolsep}{0.5pc} 
\renewcommand{\arraystretch}{1.4} 
\begin{tabular}{cccccccc}  
\hline
\multicolumn{7}{c}{\hspace{3.0cm} RMF Parametrizations} \\
 Nucleus   &  Gap    & NL1    & NLZ   & NL3   & NL3*  & NLSH    \\ \hline 
$^{96}$Cd   &  Z=48   & {\bf 1.75}   & 1.93  & 2.43  & 2.27  & 2.71   \\
           &  N=48   & {\bf 2.00}  & 2.07  & 2.59  & 2.44  & 3.03   \\ \hline
$^{108}$Cd  &  Z=48   & {\bf 1.62}   & 1.66  & 2.23  & 1.99  & 2.06   \\   
           &  N=59   & 1.30  & 1.70  & 1.50  & 1.46  & {\bf 1.20}    \\ 
           &  N=61   & 1.20   & {\bf 0.74}  & 1.20  & 1.19  & 1.50   \\ 
           &  59+61  & 2.50   & {\bf 2.44}  & 2.70  & 2.65  & 2.70   \\ \hline
$^{107}$Cd  &  Z=48   & {\bf 1.70}   & 1.73  & 2.22  & 2.18  & 2.27   \\
           &  N=59   & 1.89   & 2.16  & 2.08  & 2.04  & {\bf 1.74}   \\ \hline
$^{109}$Cd  &  Z=48   & {\bf 1.52}   & 1.61  & 1.89  & 1.84  & 1.54  \\
           &   N=61  & 1.37   & {\bf 1.16}  & 1.83  & 1.74 & 2.16  \\ \hline 
\end{tabular}\\[2pt]
\label{Table2}
\end{table}

   Two factors make the observation of discrete HD bands in $^{108}$Cd 
\footnote{Two bands with very extended shapes observed in $^{108}$Cd in
Refs.\ \cite{Cd108-1,Cd108-2} were assigned as superdeformed in Ref.\ \cite{Cd108}.}
with existing facilities less probable than in odd-mass $^{107,109}$Cd nuclei.
First, the yrast HD line in this nucleus is built from two signature 
degenerate configurations (Fig.\ \ref{DOS-HD-odd}b) in which the last neutron 
is placed into one of the signatures of the $[413]7/2$ orbital 
(see Fig.\ \ref{Cd107-both} and Ref.\ \cite{Cd108}). This reduces the feeding 
intensity of each of these bands by factor of 2 as compared with the case when 
the yrast HD line is built from single configuration. Second, the energy gap 
between the yrast and excited HD configurations decreases with increasing spin 
(Fig.\ \ref{DOS-HD}b). As a result, further reduction of feeding intensity of 
the yrast HD bands is expected if the bands are populated at spins higher than 
the spin at which they become yrast. On the contrary, the energy gap between 
the yrast and excited HD configurations is more constant as a function of spin 
in $^{109}$Cd and especially in $^{107}$Cd. {\it  All these results strongly 
suggest that the $^{107}$Cd nucleus is the best candidate for the experimental 
search of the discrete HD bands.} This conclusion is also supported by detailed 
analysis of the single-particle routhians in the yrast HD configurations of 
even-even nuclei  studied in Ref.\ \cite{CRMF-HD}; this analysis does not 
suggest any alternative case which would provide similar or larger gap 
between the yrast and excited HD configurations in even-even, odd and 
odd-odd nuclei of the $Z=40-58$ part of the nuclear chart.

      The calculated properties of the yrast HD bands in studied nuclei are shown
in Fig.\ \ref{Calc-quant}. The HD shapes undergo a centrifugal stretching that result
in an increase of the transition quadrupole moments $Q_t$ with increasing rotational
frequency. This process also reveals itself in the dynamic moments of inertia: they
increase with increasing rotational frequency in the frequency range of interest. 
On the other hand, the mass hexadecapole moments $Q_{40}$ do not show a clear trend
as a function of rotational frequency and stay nearly constant in the majority of 
the HD bands. Unpaired band crossings due to interaction of different single-particle 
orbitals are seen in the configurations of the yrast HD bands in $^{100,102,106}$Cd 
nuclei. For example, the interaction between the $(r=+i)$ signatures of the $\nu [770]1/2$ 
and $\nu [532]5/2$ orbitals is responsible for the crossing seen at $\Omega_x \sim 1.05$ MeV in 
the yrast HD band in $^{106}$Cd. This crossing may be an extra factor (in addition to 
the density of the near-yrast HD bands) which complicates the observation of the HD 
bands in $^{106}$Cd: such bands have not been observed in experiment of Ref.\ 
\cite{F.05}.

   The current study clearly shows that the polarization effects in time-even and 
time-odd mean fields have an important impact on the density of the HD states and 
especially on the energy gap between the yrast and excited HD states. The latter 
quantity is appreciably smaller (by up to $\sim 0.5$ MeV; compare Figs.\ \ref{DOS-HD} 
and \ref{DOS-HD-odd} with Table \ref{Table2}) than the respective HD shell gap in the 
routhian diagram. 

   The role of time-odd mean fields in the definition of the energy gap between 
the yrast and excited HD configurations is quite complicated. This is illustrated 
by the fact that the energy gap between the yrast HD and the lowest excited 
proton and neutron HD configurations is larger by $\approx 0.2$ MeV in the 
calculations without NM than in the ones with NM at spins where the HD 
configurations become yrast $(I\approx 67\hbar)$. This fact reflects two 
different mechanisms by which the time-odd mean fields affect the relative 
energies of different rotational bands. In the first mechanism, the angular 
momentum content of the single-particle orbitals is modified in the presence of 
time-odd mean fields, see Ref.\ \cite{AR.00} for details. There are two important 
consequences of this mechanism. First, the same total angular momentum 
of the system is built at rotational frequency which is by $\sim 25\%$ lower in the 
calculations with NM than in the calculations without NM. Second, the changes of the 
single-particle angular momenta of the single-particle orbitals surrounding the HD gaps 
of interest (the $\pi [420]1/2$ and $\pi [541]1/2$ orbitals for proton subsystem and 
$\nu [413]7/2$ and $\nu [651]3/2$ for neutron subsystem (Fig.\ \ref{Cd107-both}))
induced by NM modify the single-particle energies of these orbitals. As a 
result, these gaps are smaller by $\sim 0.12$ MeV in the calculations with NM at 
$I=67\hbar$. The second mechanism is related to additional binding due to time-odd 
mean fields. The time-odd mean fields are stronger in the excited HD configuration 
than in the yrast HD configuration. Thus, additional binding due to NM is stronger in 
excited HD configuration than in the yrast HD configuration. This also leads to the 
decrease of the energy gap between the yrast and excited HD  configurations in the 
calculations with NM as compared with the ones without NM.
 
 The presence of time-odd mean fields reveals itself also in the energy splitting of  
the opposite signatures of the $\nu [770]1/2$ orbital visible at $\Omega_x=0.0$ 
MeV (Fig.\ \ref{Cd107-both}); the occupied orbital is more bound than unoccupied 
one in the RMF theory (Ref.\ \cite{A150}). 

  When considering theoretical predictions one has to keep in mind that they 
are subject of the errors in the description of the energies of the single-particle 
states, which exist in the RMF theory at spherical shape \cite{RBRMG.98},
normal deformation \cite{A250} and quite likely at superdeformation \cite{Cd108}.
The extrapolation from spherical and normal deformation towards HD is itself a
potential source of errors since it is not know how well the response of the mean 
field (or the single-particle potential and liquid drop in the MM method) to the 
extreme elongation of the nucleus is reproduced in model calculations. Such errors 
are not restricted to the self-consistent models; they are also expected in the 
phenomenogical potentials (used in the MM method) which describe single-particle 
energies at normal deformation better than self-consistent models. However, 
several facts support the results and interpretations given above. First, all RMF 
parametrizations used in this study lead to the same HD configurations in $^{96,107-109}$Cd
nuclei which become yrast at similar spins (see Ref.\ \cite{CRMF-HD} for comparison of the 
results obtained with NL1 and NL3) and to similar sizes of the proton and neutron HD 
shell gaps (Table \ref{Table2}). Second, the large size of the $Z=48$ and $N=59$ (and 
especially of combined neutron $59+61$ gap) HD shell gaps reduces the importance of the errors 
in the description of the energies of specific single-particle states. Third, the MM 
results of Ref.\ \cite{A100} suggest similar conclusions for the nuclei around 
$^{108}$Cd. Indeed, large $Z=48$ shell gap and low density of the single-particle states 
in the vicinity of the $N=59$ and $N=61$ HD shell gaps is clearly visible in Figs. 4 and 
5 of Ref.\ \cite{A100}. The $N=59$ and $N=61$ shell gaps are separated by the 
signature-degenerated $7/2^+$ state (Fig. 5 in Ref.\ \cite{A100}). Thus, similar to our 
case, the yrast HD line in $^{108}$Cd will be formed from two signature degenerated 
configurations in the MM calculations.

\section{Conclusions}

  In summary, a systematic analysis of hyperdeformation in the Cd isotopes 
has been performed in the cranked relativistic mean field theory. The density
of the HD states has been analysed with the goal to find the best cases for 
experimental search of the discrete HD bands. Our analysis indicates $^{96}$Cd 
as a doubly magic HD nucleus in this part of nuclear chart; its magicity is 
due to large $Z=48$ and $N=48$ HD shell gaps. However, experimental study of 
HD in this  nucleus is problematic with existing facilities due to its $N=Z$
status. The low density of the neutron single-particle states in the vicinity 
of the $N=59$ and 61 HD shell gaps and sizable $Z=48$ HD shell gap lead to 
appreciable gaps between the yrast and excited HD bands in $^{107-109}$Cd nuclei, 
thus offering better opportunities to observe discrete HD bands. Among these 
three nuclei, the best candidate for observing the discrete HD bands with 
existing facilities is $^{107}$Cd nucleus. The MM calculations of Refs.\ 
\cite{A100,SDH.07} indicate that the fission barriers are sufficiently large 
in the nuclei around $^{108}$Cd so that the HD minimum could survive fission 
for a significant range of angular momentum.  The stability of the HD 
minimum is defined by its depth, the fission barrier height and the height of 
the barrier between the HD and normal-deformed/superdeformed minima 
\cite{Dudek,SDH.07}. Our study clearly indicates that the HD minimum is 
localized in the potential energy surface. However, future studies of the 
HD in this mass region have to provide more quantitative answers on these
properties of the HD minima in a fully self-consistent framework.

  The work was supported by the U.S. Department of Energy under grant 
DE-FG02-07ER41459.

\end{document}